\documentclass[11pt]{article}
\usepackage[margin=80pt]{geometry}

\usepackage[utf8]{inputenc} 
\usepackage[T1]{fontenc}    
\usepackage{hyperref}       
\usepackage{url}            
\usepackage{booktabs}       
\usepackage{amsfonts}       
\usepackage{nicefrac}       
\usepackage{microtype}      
\usepackage{lipsum}    
\usepackage{graphicx}
\usepackage{amsmath}

\usepackage[title]{appendix}
\usepackage{hyperref}
\usepackage{tabu}
\usepackage{multirow}
\usepackage{longtable}
\usepackage{graphicx}
\usepackage{amsmath}
\usepackage{appendix}
\usepackage{pgfplots}
\usepackage{array}
\usepackage{url}
\usepackage{color,soul}
\usepackage{enumerate}
\usepackage[flushleft]{threeparttable}
\usepackage{tikz}
\usetikzlibrary{patterns}
\usepackage{color}
\hypersetup{
    colorlinks,
    citecolor=black,
    filecolor=black,
    linkcolor=black,
    urlcolor=black
}
\title{Geographical Security Questions for Fallback Authentication}

\author{
  Alaadin Addas, Julie Thorpe, and Amirali Salehi-Abari\\
  Faculty of Business and IT\\
  University of Ontario Institute of Technology\\
  \textit{\{alaadin.addas, julie.thorpe, abari\}@uoit.ca}\\
}  

\date{}
\begin{document}
\maketitle

\begin{abstract}
Fallback authentication is the backup authentication method used when the primary authentication method (e.g., passwords, fingerprints, etc.) fails. Currently, widely-deployed fallback authentication methods (e.g., security questions, email resets, and SMS resets) suffer from documented security and usability flaws that threaten the security of accounts. These flaws motivate us to design and study Geographical Security Questions (GeoSQ), a system for fallback authentication. GeoSQ is an Android application that utilizes autobiographical location data for fallback authentication. We performed security and usability analyses of GeoSQ through an in-person two-session lab study (n=36, $18$ pairs). Our results indicate that GeoSQ exceeds the security of its counterparts, while its usability (specifically login time) has room for improvement. 
\end{abstract}


\section{Introduction}
\label{sec:Introduction}

Authentication mechanisms (e.g., passwords, biometrics, PINs, etc.) play a critical role in securing our accounts and devices against unwanted access and intrusions. However, our primary means of authentication fail when we forget our \emph{secrets} (e.g., passwords or PIN) or when our biometric measurement is malfunctioning (e.g., from a cut on your finger tip). These failures of authentication mechanisms motivate the urgency of backup/ secondary authentication, usually referred to as \emph{fallback authentication}, for the users to gain access to their accounts or devices. 

The most popular fallback authentication methods are security questions, email resets, and SMS resets. Security questions (or personal knowledge questions) are often in the form of predefined questions (e.g., what is the color of your first car?). For fallback authentication, the users' answers to these questions must match his/her answers provided at registration time. Some other popular fallback authentication methods use other communication channels such as email or phone to send a link or PIN for password reset. These fallback authentication methods suffer from security flaws that compromise the security of our accounts and devices.

Security questions are easy to guess by an adversary \cite{Golla2013}. Email resets rely on the security of the email account for all other accounts. This makes the email account a point of attack because if the email is breached then so are many other accounts (avalanche effect) \cite{Garfinkel2003}. Email resets also exhibit some usability problems specifically when users don't have access anymore to their preset recovery email \cite{Garfinkel2003}. SMS resets have similar issues, and also are prone to attacks on telecommunication protocols \cite{Welch2017}. These three fallback authentication methods (i.e., security questions, email resets, and SMS resets) are also vulnerable to the known adversary attack. The known adversary is any individual with first-hand knowledge of a potential victim and/or elevated access to a potential victim’s devices, who uses these privileges with malicious intent. The known adversary can easily guess the answer to the security questions or gain access to a potential victim's devices in order to initiate the fallback authentication process such as an email or SMS reset. 

These security flaws in widely-utilized fallback authentication methods motivate us to explore alternative fallback authentication methods. We specifically investigate the usability and security of GeoSQ (Geographic Security Questions) as a means of fallback authentication. In GeoSQ, users are expected to answer a sequence of  autobiographical location questions (e.g., where were you on the 18th of December at 4:00 PM?) by clicking on a digital map.  GeoSQ attempts to address some of the aforementioned security flaws such as the easy guessability of security questions, the avalanche effect vulnerability in email resets and SMS resets, and attacks on telecommunications protocols.

We investigate the security and usability of GeoSQ through a user study that spanned two sessions over 7-11 days with 36 participants.

From a security perspective, our results indicate that GeoSQ is resilient to throttled online guessing attacks, and phishing attacks. However, GeoSQ is not resilient to the known adversary threat due to the predictability of locations by known adversaries. The large key space of $2^{94.25}$, offered by GeoSQ,  makes it very difficult to conduct a successful throttled online guessing attack (see Section \ref{subsec:securityanalysis}). When compared to the security of security questions, email resets, and SMS resets, GeoSQ offers improved protection against several threats including phishing attacks, throttled online guessing attacks, and unthrottled guessing attacks (see Appendix \ref{appendA} for a comparison).

From a usability perspective, GeoSQ needs improvement in several key metrics. The average login time for participants is about $5$ minutes and $50\%$ of questions were answered correctly. The long login time and the frequency of errors is a point of concern when compared to currently utilized fallback authentication. Security questions have a high failure rate of $39$\% \cite{Golla2013}, while email resets and SMS resets have lower failure rates ($25$\% and $19$\% respectively) \cite{Bonneau2015, Golla2019}. These three authentication techniques also offer quicker login times ($34$ seconds for email resets, $98$ seconds for SMS resets, and $16$ minutes for security questions \cite{Golla2019}.  

Our study and investigation has shed light on important future work to make GeoSQ more usable while still maintaining its security. To address memorability concerns in GeoSQ, one promising direction is to detect significant personalized events for each user, and tailor location security questions towards such events. However, the selection of such events needs careful investigation as those events need to be memorable yet not easily guessable by an adversary. By addressing the remaining usability issues of GeoSQ, it may be a viable alternative to widely-deployed fallback methods which suffer from documented security and usability flaws.

\section{Related Work}
\label{sec:relatedwork}
The most popular methods for fallback authentication are security questions, email resets, and SMS resets. Autobiographical authentication has also recently attracted attention as a viable alternative \cite{Das2013, Hang2015c, Albayram2016}. We first review the literature on these methods while highlighting some of their current security and usability shortcomings. We then review some primary knowledge-based authentication frameworks and research which influence the design and development of GeoSQ.

\vskip 5mm
\noindent \textbf{Security Questions}. In this class of fallback authentication, also known as personal knowledge questions, users are expected to recall their preset answers to security questions that set by a verifier or themselves. Just \emph{et al.} \cite{Just2009} found that answers to security questions are low in entropy, and consequently easy-to-guess, despite verifiers' attempts in guiding users in creating more difficult-to-guess answers to security questions. Evaluating a leaked set of $3.9$ million answers to security questions, Golla \emph{et al.} \cite{Golla2013} confirm similar results where users' answers were easy to guess.

Bonneau \emph{et al.} \cite{Bonneau2015} performed usability and security analyses of security questions. They discovered that $40$\% of users failed their password recovery attempts through security questions, and $37$\% users attempted to make their answers less guessable by adding a word to the beginning or end of the answers. This often had the opposite security effect as noted by Golla \emph{et al.} \cite{Golla2013}.

\vskip 5mm
\noindent \textbf{Email Resets.} Email resets rely on a pre-enrolled email through which a verifier sends a reset link or a verification code for password reset. Garfinkel \cite{Garfinkel2003} studied the usability and security of email resets. Email resets make the email a single point of attack, because if the recovery email is compromised many other accounts are easily compromised. Email resets suffers from a usability issue with the loss of the recovery email, as this loss would complicate the fallback authentication process \cite{Garfinkel2003}.

Guri \emph{et al.} \cite{Guri2016} performed an analysis on account recovery techniques for popular social networking and email platforms such as Facebook, Gmail, and Twitter. It is shown that rogue applications in a mobile environment can request access to sensitive resources such as email, thus giving an attacker the capability of snooping on email resets.

\vskip 5mm
\noindent \textbf{SMS Resets.} SMS resets rely on a pre-enrolled phone number in order to send a reset link or a verification code to the user.  It haa been shown that the channel of communication (cellular network) for SMS resets is susceptible to snooping attacks \cite{Al-Maqlabi2018,Lilly2017} and flaws in telecommunication protocols \cite{Welch2017}.  Snooping attacks via the utilization of an IMSI catcher can compromise the contents of an SMS message \cite{Lilly2017, Al-Maqlabi2018}.\footnote{An IMSI catcher is also known as a \emph{Stingray} is a device that mimics a cell tower and sends signals to target phones to trick them into connecting to it  and sharing information that should only be shared with a cell tower.} Also, flaws in telecommunications protocols such as SS7---a protocol for letting telecommunication companies share billing information---allows attackers to intercept call and SMS data \cite{Welch2017}. Publicly available tools \cite{SS7} can be utilized to exploit flaws in the SS7 protocol. Both IMSI catchers and flaws in telecommunication protocols put SMS resets at risk of interception. In addition to these vulnerabilities, Guri \emph{et al.} \cite{Guri2016} noted that rogue applications can request access to SMS. Thus, SMS reset codes or even reset links can easily be comprised using such rogue applications in a mobile environment.

\vskip 5mm
\noindent \textbf{Autobiographical Authentication.} The aforementioned security and usability flaws have led the investigation of alternative authentication techniques. Das \emph{et al.} \cite{Das2013} performed a large study on autobiographical authentication. Utilizing an MTurk study (n=$70$),  $9$ distinct categories of autobiographical data were determined including location data from everyday activities. Another online study (n=$145$) was run to assess the memorability of autobiographical data by assigning memorability score to each autobiographical data type. The findings of these two studies helped the development of \emph{MyAuth}, an application that logs different types of autobiographical data and queries the user about them. Through a field study (n=$24$), with $2147$ autobiographical questions asked, $1381$ (~$64$\%) were answered correctly. It was found that location questions of ``where were you on <time>'' were more likely to be answered correctly than app usage questions of ``What application did you use on <time>.'' This was part of our motivation to further investigate autobiographical location in GeoSQ.

Hang \emph{et al.} \cite{Hang2015c} conducted research on $7$ categories of autobiographical authentication data. Their pre-study (n=$19$) showed that outgoing SMS, incoming SMS, and app usage questions are the most promising in terms of memorability and repelling adversaries (accuracy = $92.8$\%, $79$\%, and $70$\% respectively). Accuracy scores incorporate the total number of correct answers from legitimate and incorrect answer from adversarial users over the total number of login attempts. Recruited adversaries (n=$19$) were highly successful in guessing outgoing SMS and incoming SMS questions ($65$\%, and $61.4$\% respectively); however, adversaries were not highly successful in guessing app usage ($35$\%). The pre-study was followed up with a main study ($n=11$) participants were asked to bring along two adversaries (one socially close adversary and one acquainted adversary). App usage had the strongest accuracy rating, followed by incoming and outgoing calls ($81.8$ \%, $72.7$\%, $72.7$\% respectively). 

AlBayram \emph{et al.} \cite{Albayram2016} conducted a field study (n=$24$) on $9$ categories of autobiographical authentication data.  A monitoring application was utilized to log autobiographical data, and participants were asked questions  from the previous $24$ hours for each category. It was noted that autobiographical data has episodic memory, which tends to be more memorable in shorter time spans \cite{Conway2009}. The most promising categories of autobiographical data in terms of memorability were incoming/outgoing call, and location data ($0.76$, and $0.69$ respectively). As with Hang \emph{et al.} \cite{Hang2015c}, two types of adversaries (strong vs. naive) were recruited to guess autobiographical data. The adversaries obtained very low accuracy scores for incoming/outgoing calls (strong adversary accuracy: $0.13$, naive adversary accuracy: $0.0008$). The accuracy scores for location questions were $0.28$ for strong adversaries and $0.038$ for naive adversaries. 

Both Hang \emph{et al.} \cite{Hang2015c} and AlBayram \emph{et al.} \cite{Albayram2016} investigate the threat of social insiders on their proposed autobiographical authentication systems. Muslukhov \emph{et al.} \cite{Muslukhov2013} empirically discerned the frequency of unauthorised access to smartphones by insiders through a user study. We refer to insiders as known adversaries to avoid confusion with the common utilization of the term which refers to the threat of insiders within an organization \cite{Warkentin2009}.

This related work provides support that autobiographical data can be effectively utilized to authenticate users. We developed GeoSQ with a focus on location-based autobiographical authentication, based on the promising findings in the studies conducted by Das \emph{et al.} and AlBayram \emph{et al.} \cite{Das2013, Albayram2016}. The memorability of location-based autobiographical questions was usable \cite{Das2013, Albayram2016}. We added some features to our system to ensure enhanced security. Furthermore, when evaluating GeoSQ for security, we ensured that we tested against the known adversary threat by recruiting participants in pairs, because it was identified as a pertinent threat \cite{Muslukhov2013, Hang2015c, Albayram2016, Das2013}.

\vskip 5mm

\noindent \textbf{Other Relevant Authentication Systems.} Alphanumerical passwords are the most widely-used method for primary authentication. However, recent research has detailed their usability and security weaknesses  \cite{Bonneau2010}. Modern password crackers have been proven to be effective in efficiently guessing a large number of passwords. Notable example of password crackers use various artificial intelligence techniques such as Probabilistic Context-Free Grammar (PCFG) \cite{Veras2014, Weir2009}, Ordered Markov Enumerators (OME) \cite{Durmuth2015}, and Artificial Neural Networks (ANNs) \cite{Melicher2016b}. These recent development in password cracking has motivated the investigation of alternative primary authentication systems.

Graphical passwords have drawn considerable attention as an alternative for alphanumerical passwords \cite{Biddle2012}. A well-studied class of graphical password is click-based graphical passwords (e.g., PassPoints \cite{Wiedenbeck2005}, Cued Click Points \cite{Chiasson2007}, and Persuasive Cued Click Points \cite{Chiasson2008b}). These systems expect the users to choose and recall a sequence of points on a set of background images as their passwords. For usability issues, these systems authenticated users if their click-points have acceptable \emph{error margin} to their selected points. Error margins are implemented through the process of discretization \cite{Birget2003, Chiasson2008, Suo2005}. The security of various passpoint-style graphical passwords is studied  \cite{Thorpe2008,Thorpe2007,Thorpe2014a,Zhao2015} which has motivated the development and design click-based authentication systems on videos \cite{Thorpe2012}, and digital maps \cite{Hang2015b, Thorpe2013, MacRae2016}. Among these systems, map-based authentication systems (e.g., GeoPass and GeoPassNotes \cite{Thorpe2013, MacRae2016}), are of relevance to our work. GeoPass authenticates a user by setting a marker on a map, and GeoPassNotes combines map-based authentication with text based authentication by authentication users through a combination of a note and a marker on a map.

The memorability of geographical authentication systems that rely on digital maps like GeoPass and GeoPassNotes is very high ($97$\% and $100$\% respectively) after $1$ week of setting the credentials. The GeoPass results were validated in a separate study by Al-Ameen and Wright \cite{Al-Ameen2014}. Al-Ameen and Wright conducted a second study on GeoPass to determine the effects of multiple password interference. The results shows that $\sim70$\% of users were able to successfully authenticate, despite the multiple password interference effect \cite{Al-Ameen2015}. Subsequently, a third study was conducted on GeoPass where participants were asked to create a mental story in the credential setting phase \cite{Al-Ameen2015}. The results showed that creating a mental story reduced failures due to interference effects to $3.7\%$ \cite{Al-Ameen2015}.

\section{GeoSQ: Implementation and Design Decisions}
\label{sec:GeoSQImplementationandDesignDecisions}
We designed and developed a location-based fallback authentication system called Geographical Security Questions (GeoSQ), a variant of previously-proposed autobiographical authentication systems \cite{Albayram2016, Hang2015a}.\footnote{GeoSQ is an Android application developed using Google Maps API \cite{GoogleMapsAPI}.} We analyse the security and usability of GeoSQ through an in-person user study (see Section \ref{sec:user_study} and Section \ref{sec:results} for details).

\begin{figure}[tb]
\begin{tabular}{cccc}
 \includegraphics[width=0.225\textwidth]{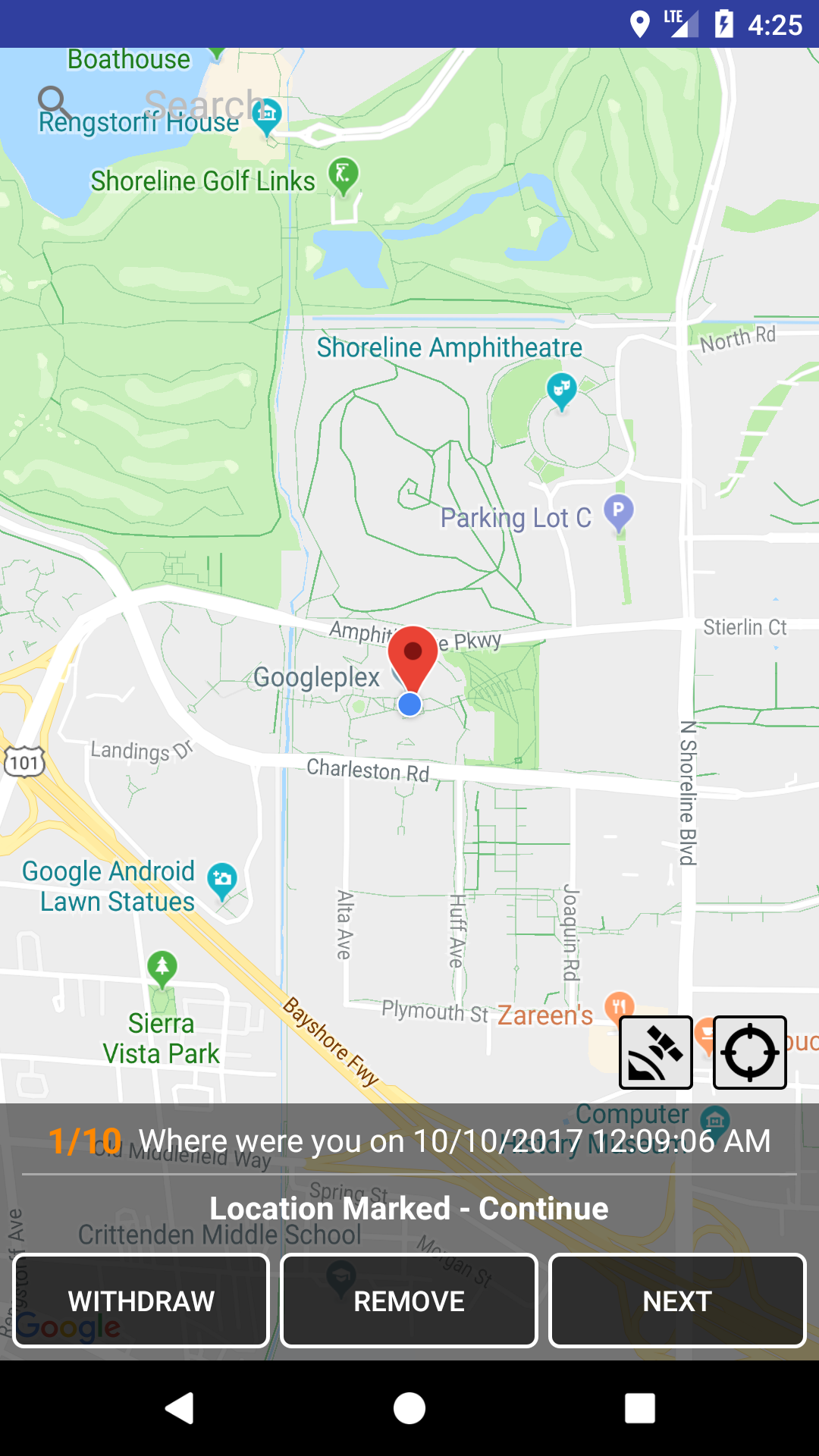} &  \includegraphics[width=0.225\textwidth]{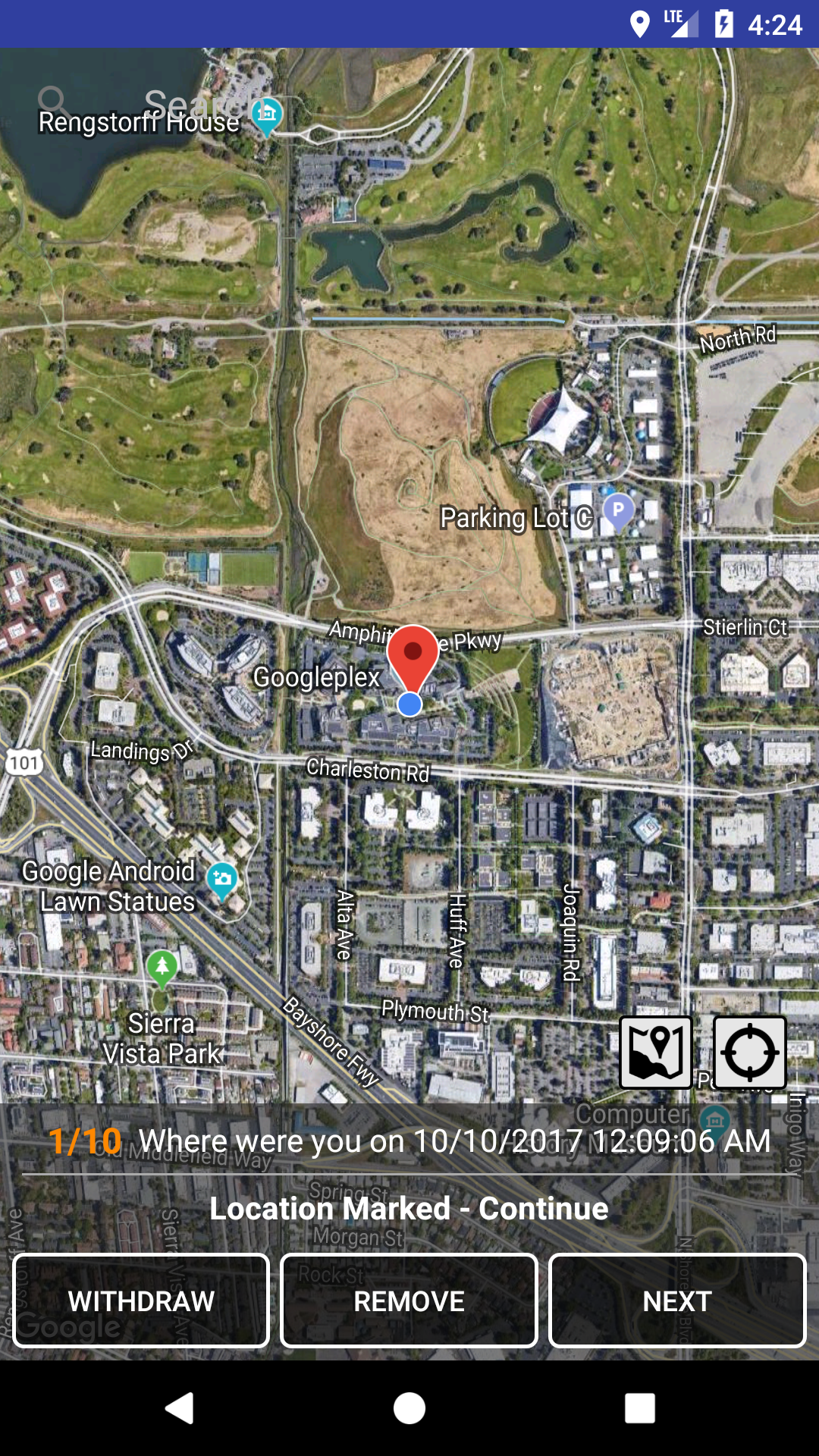} &  \includegraphics[width=0.225\textwidth]{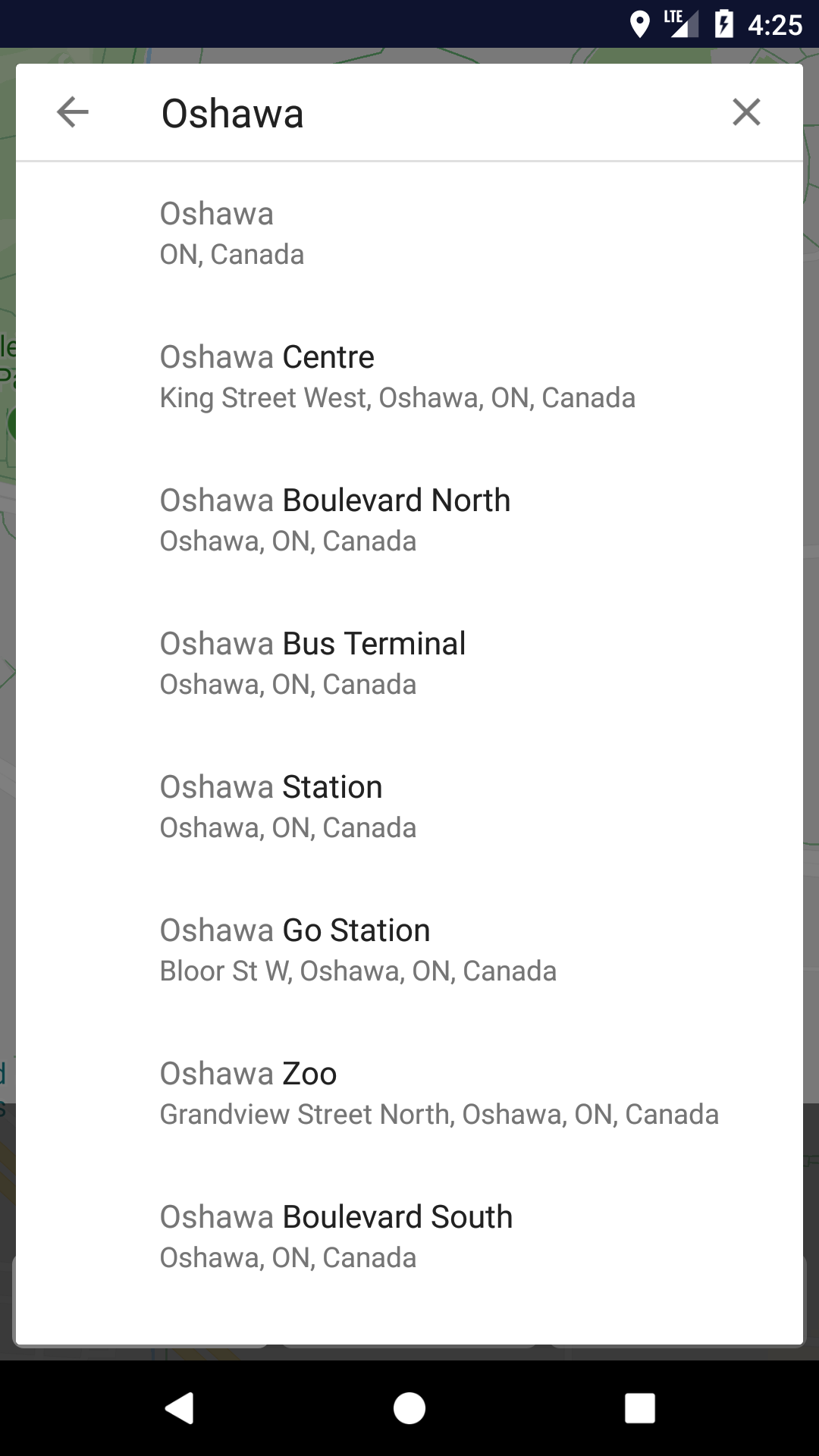} &  \includegraphics[width=0.225\textwidth]{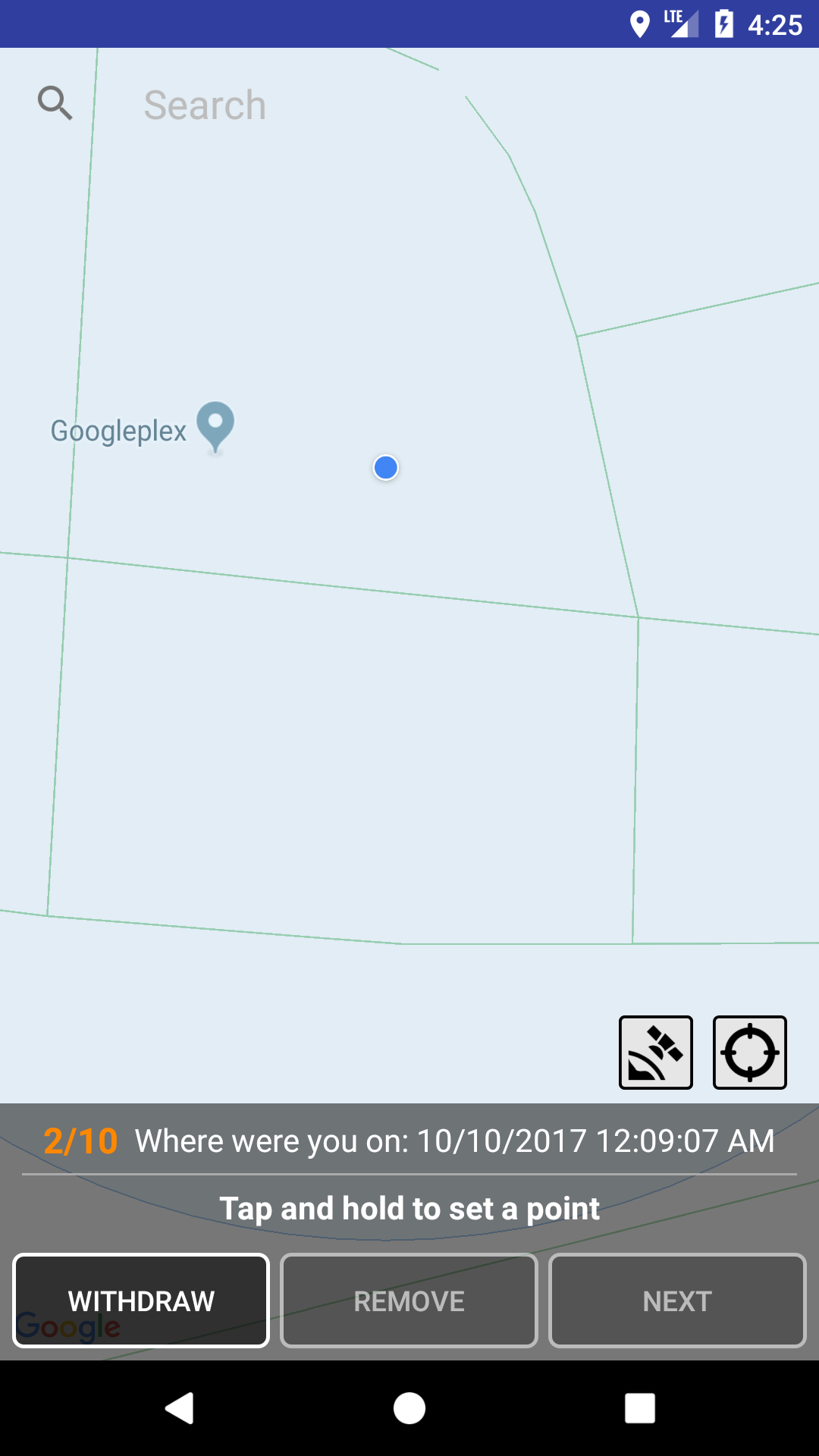}  \\
(a) Default Mode & (b) Satellite Mode & (c) Search  & (d) Zoom   \\
\end{tabular}

\caption{GeoSQ Interface; (a) Default map mode, users can set/remove markers and navigate to current location; (b) Satellite map mode that users can switch to for better memorability; (c) Search functionality for easy navigation; and (d) Zoom functionality for fine grain location setting.}

\label{fig:GeoSQ-interface}
\end{figure}

GeoSQ is designed to run in the background with enabled location services to log unique locations visited by the user. A location is considered \emph{visited} if the user has stayed at least 5 minutes in that location. The \emph{uniqueness} of locations is determined by checking if a location is $400$ meters away from any previously logged location.\footnote{A 400 meter threshold was set to ensure that participants on a campus or a large building do not obtain multiple location questions in the same vicinity, we utilized our own campus to generate this number as all of our participants were affiliated with our university.} Locations are logged in GeoSQ by geographic coordinates (latitude and longitude). When 10 locations have been logged, the user can be queried about their unique visited locations in the following format: Where were you on the $dd$ of $mm$ at $t$, where $dd$, $mm$, and $t$ stands for specific day, month, and time, respectively (e.g., where were you on the $14^{th}$ of February at 4:00PM?). The user is then expected to navigate to a location on the map and set a marker on the correct logged location. A response location is correct if the marker is set within $200$ meters of the logged location. For a successful authentication, the user must answer 7 out of 10 location questions correctly. As shown in Fig.~\ref{fig:GeoSQ-interface}a, the user is required to click the next button after answering a location question (i.e., dropping a marker on the location). The user can change the selected location by using the remove button to first deselect the location, and then dropping a marker on the new location. Lastly, users have the ability to withdraw at any time and uninstall the application by clicking on the withdraw button.

Users can also switch between default map mode (see Fig.~\ref{fig:GeoSQ-interface}a) and satellite map mode (see Fig.~\ref{fig:GeoSQ-interface}b), provided by the Google Maps API. Map navigation can be done either by dragging over the map, or using the search bar with location keywords (see Fig.~\ref{fig:GeoSQ-interface}c). Search results are ordered based on the current location. For example, when searching for a popular restaurant chain, the branches closest to the user's current location will be displayed first. GeoSQ users also have the ability to zoom in and out (see Fig.~\ref{fig:GeoSQ-interface}d).

GeoSQ was implemented with several usability and security goals in mind. When designing GeoSQ, we also incorporated the findings of other related research in autobiographical authentication \cite{Hang2015c, Albayram2016, Albayram2015}. We explain our security-oriented and usability-oriented decisions below.

\subsection{Security-Oriented Design Decisions}
An important security concern with location-based autobiographical authentication is that the daily mobility patterns of users are predictable (e.g., users go to work and return home during weekdays). This makes mounting attacks easier (even  with limited guesses). To address this security concern, we filter out these predictable easy-to-know locations by a simple heuristic. We assume that the locations at which the user spent more than 5 hours are predictable and easy-to-know by adversary. This security decision has a usability cost, as GeoSQ requires a longer period of time (e.g., $7$--$10$ days) to log enough unique, and less predictable locations.

To ensure that GeoSQ has reasonable resilience to guessing attacks, GeoSQ asks a user 10 unique location questions, and requires 7 (out of 10) correct answers for a successful authentication. The choice of $70\%$ threshold is supported by earlier findings that users are able to recall roughly $70\%$ of their locations \cite{Albayram2016}. Our decision to ask the users 10 questions was to ensure a key space that can withstand an offline guessing attack (see Section \ref{subsec:securityanalysis} for analyses). However, requiring 10 questions limits the usability of GeoSQ where users should remember locations visited a few days ago, and users' memorability of locations plummet after 24 hours \cite{Albayram2016}.

\subsection{Usability-Oriented Design Decisions}
When utilizing a classical alphanumeric passwords, exact match between selected and entered password is necessary for a successful authentication. However, it is unrealistic to expect users to achieve that type of accuracy for location-based autobiographical authentication systems. As such, we considered a 200-meter error margin that would account for human input errors as well as errors in the accuracy of logging locations. This hinders security because it makes the key space smaller, but is necessary for usability purposes. Our decision on 200-meter error is based on \emph{location accuracy settings} (discussed below) and \emph{input errors} in touch screen interfaces. Through our pilot studies, we noticed that input errors vary depending on the zoom level. Therefore, we set the default zoom level to be $16$, at that zoom level, the 200-meter error margin is effective enough to reduce such errors.

Location services in smartphones drain a considerable amount of battery power. Thus, we adjusted the location setting so as to minimize the risk of missing locations and to keep the battery usage low. So we set GeoSQ to refresh the location every $2.5$ minutes with the \emph{balanced power} setting.\footnote{There are three location accuracy settings that Google Maps API \cite{GoogleMapsAPI} offers: \emph{high accuracy} which is the most accurate location request but utilizes a great amount of battery power; \emph{balanced power} provides great accuracy but does not consume as much battery power as high accuracy; and \emph{low power} is not very accurate but consumes very little battery power.} This location setting provides us an accuracy with an error margin of a city-block \cite{GoogleMapsAPI}, which is around 200 meters on average.\footnote{Related work utilized a 75-meter margin of error based on the Haversine distance \cite{Albayram2016, Albayram2015b}. The Haversine distance is the great circle distance between two points on a sphere given their latitudes and longitudes. Our greater error tolerance (i.e., 200 meters) is based on the location accuracy of balanced power setting in  Google Maps API \cite{GoogleMapsAPI}.} 

We emphasize that our decision to log visited locations, at which user remained for more than 5 minutes, was necessary to prevent logging transient locations (e.g., a sequence of locations where a user is walking from work to his vehicle). Recalling transient locations are hard especially over a span of 7-11 days.

\section{User Study and its Design}
\label{sec:user_study}

We evaluated the security and usability of GeoSQ through a 38-participant (19 pairs) user study, approved by our university’s Research Ethics Board. Before the user study, GeoSQ was pilot tested by 6 volunteers. Three pilot testers were experienced computer users with degrees in computer science or IT, the remaining three were casual computer users. Our pilot testing allowed us to discover and remove usability flaws and bugs within our system, and improve the user study's instructions.

Participants for the user study were recruited using a broadcast email and posters on campus. Participation was limited to students, visitors, and staff of our university. All participants must have met the following criteria: (i) 18 years of age or above; (ii) Participants must bring a pair; (iii) Participants must have an Android smartphone; and (iv) Participants must be willing and able to turn on location services throughout the week.

Our user study contained two sessions spanning $7$--$11$ days. The pairs completed the exact same steps (i.e., we did not have a main participant and a pair).

\vskip 10pt
\noindent\textbf{Session 1.} This session was an in-lab session held on several different dates, and time slots (with the maximum of two pairs in each time slot). Each participant was compensated $\$8$ for their participation.  We asked pairs to sit across from each other to avoid any type of contamination of the results. Followed by reading the pre-written instructions to the participants, we also ran a demonstration of the GeoSQ application. After reading the consent form and agreeing to it, they proceeded to complete the entry survey---a mixture of background questions (e.g., age, gender, and academic background) as well as the self-declared relationship to the pair (e.g., my pairs is my close friend). We instructed participants to download, install and read the embedded instructions within GeoSQ.  Lastly, they would be reminded to keep location services on and we reminded them that GeoSQ is logging location information \emph{locally} in the background. They were informed that location services could be turned off when they were not comfortable with logging locations.

\vskip 10pt
\noindent\textbf{Session 2.} This session was held $7$--$11$ days after Session 1 in the lab to provide enough time for logging $10$ unique visited locations. Participants were compensated $\$10$ for their participation. Additionally, the participating pairs were entered into a $\$100$ draw if both returned for Session 2. We note that 36 (out of 38) participants of Session 1 returned for Session 2. 

Similar to Session 1,  Session 2 also started with the instructions being read off a pre-written script. GeoSQ prompted each participant with ten location questions regarding their whereabouts of their previous $7$--$11$ days. Then, each pair was asked to switch phones and attempt to guess each other's  location questions.\footnote{The participants were asked to switch phones because the autobiographical location data was logged locally.} A set of $10$ identical questions were asked for each user and his/her paired attacker.  We also kindly requested that the participants not communicate at all during this period to avoid the contamination of results.

During the location recall and location guessing phases, the users were actively encouraged to use the Internet for research. We observed that participants used sources such as their Google Timelines or transportation system cards to find some information regarding their whereabouts at a certain point in time. Lastly, participating pairs returned phones to each other and answered usability questions in an exit survey.

In Session 1 and Session 2, we also tested another completely independent authentication system that is not discussed in this paper. As another system was tested in our study, Session 1 was approximately 35 minutes and Session 2 was approximately 45 minutes. In both sessions, the GeoSQ memory test was performed after the other independent system.  However, as GeoSQ relies on memory of autobiographical events selected by the system (as opposed to the user), we do not believe this had a major effect.  

\vskip 10pt
\noindent \textbf{Demographics Details.} Recruited participants were all undergraduate students in the range of $18$--$30$ years old (with the average of $21.3$). Out of $38$ initial participants, $13$ were female ($34.2\%$), 25 were male ($65.7\%$), and 15 ($39\%$) had already taken some computer security/IT course .

\section{Security and Usability Analyses}
\label{sec:results}
We here discuss the security and usability analyses of GeoSQ. Our analyses are based on the user study discussed in Section \ref{sec:user_study}.

\subsection{Security Analysis}
\label{subsec:securityanalysis}
We analyse the security of GeoSQ under various threats that GeoSQ is expected to be resilient against as an autobiographical fallback authentication. For each threat, we first define it and then measure the resilience of GeoSQ against it.

\vskip 8pt
\noindent \textbf{Throttled Online Guessing Attacks.} Resilience to throttled online guessing attacks, formally defined by Bonneau \emph{et al.}~\cite{Bonneau2012}, is necessary for any fallback authentication method. A system is resilient to throttled online guessing attacks if an attacker cannot compromise more than $1\%$ of accounts a year, given ten guesses a day \cite{Bonneau2012}. Throttled online guessing attacks can fall into two categories based on whether or how the adversary has knowledge of victim. We here first analyze the \emph{classical} throttled online guessing attacks in which the adversary has no knowledge of the potential victim. We then analyze a special cases of \emph{known adversary} which has first-hand knowledge of the victim.

We ensured that GeoSQ is resilient to classical throttled online guessing attacks by taking two important measures. (i) We only allow one attempt per location question. This restriction effectively enhances the security of our systems at the usability cost (discussed below). (ii) We also expand the key space by requiring $7$ out of $10$ questions to be correctly answered. In our user study, we didn't track mobility patterns of our participants (due to privacy and confidentiality) as such we have estimated the key space to be $2^{94.25}$

Our key space calculation is based on the assumption that users will be within a $12$ km radius ($452.3$ $km^{2}$) of their home location. This is consistent with reported statistics for commute distances to work (the median commute is $12.9$km) \cite{StatsCanadaCommuting}.  We then incorporate the concepts of central and robust discretization \cite{Birget2003, Chiasson2008}. Given our $200$ meters error margin, one discrete location covers $0.04$ $km^{2}$. We therefore have $\frac{452.3}{0.04}= 11,307$ ($2^{13.4}$) unique locations per question. Since we require $7$ out of $10$ location questions to be answered correctly for a successful authentication, our key space is $P(11,307, 7)=2^{94.25}$, where $P(n,k)$ is the the number of k-permutations of a set with $n$ elements. As our system only allows one attempt per location question and its key space is very large, GeoSQ is resilient to throttled online guessing attacks.

\vskip 8pt
\noindent \textbf{The Known Adversary.} We differentiate between known adversary and a throttling online attacker by determining whether or not an attacker has first-hand knowledge of the potential victim.\footnote{The known adversary is not formally defined in the framework for the comparative evaluation of web authentication schemes \cite{Bonneau2012}.} Our known adversary model is similar to that of Hang \emph{et al.} \cite{Hang2015c} and AlBayram \emph{et al.} \cite{Albayram2016}. 
Consistent with the framework of Bonneau \emph{et al.} \cite{Bonneau2012}, we consider a system to be resilient to the known adversary if known-adversary attackers cannot compromise more than $1\%$ of accounts per year.
 
\begin{figure}[htbp!]
	\centering
 \includegraphics[width=0.5\textwidth]{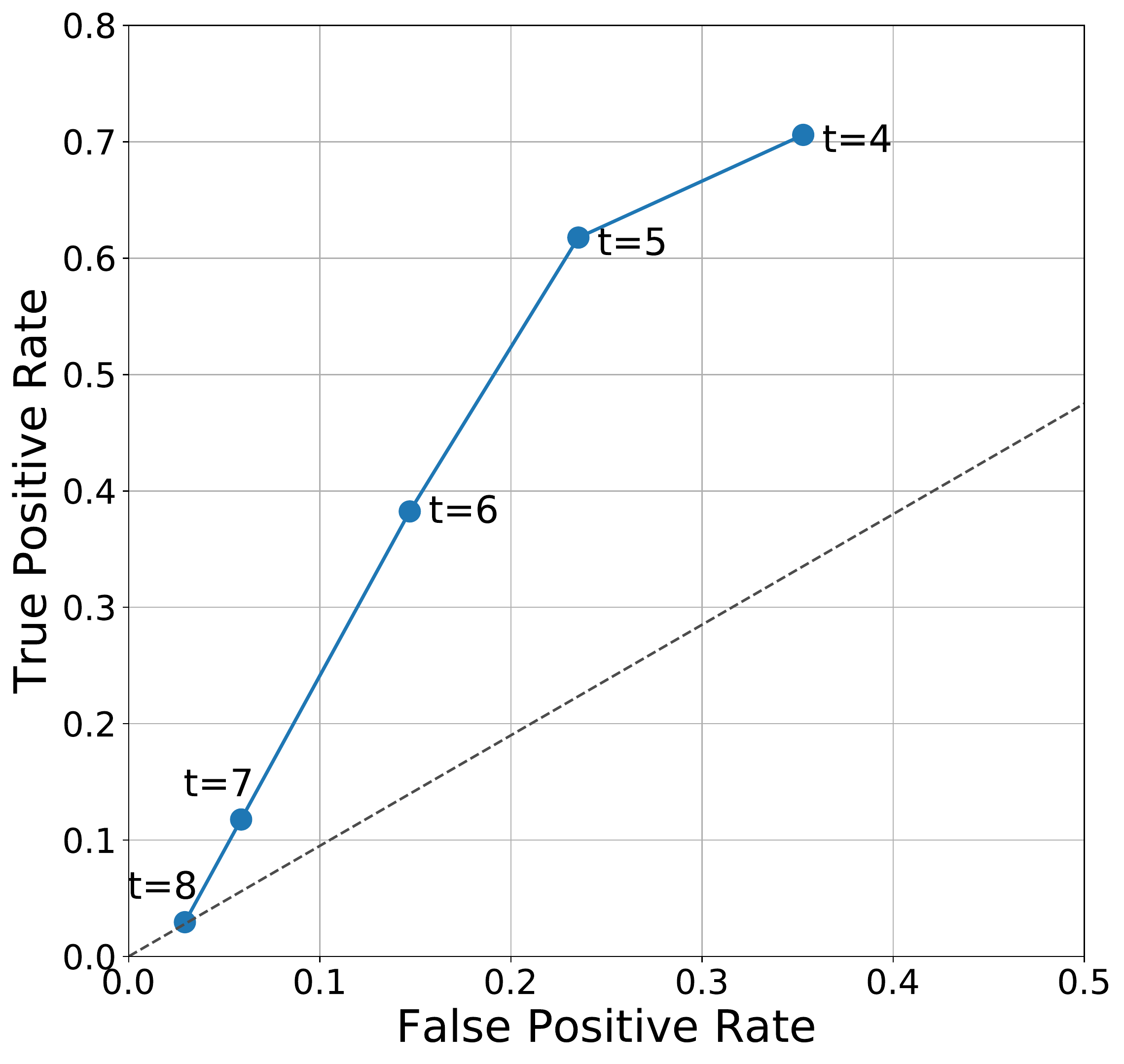}
	\caption{ROC Graph for GeoSQ with varying thresholds (t). Note that t=10 and t=9 are not shown as they both have zero true positive and false negative rates.}
	\label{fig:ROCPNG}
\end{figure}

We allowed pairs to attempt to guess each others’ autobiographical location questions. In our analysis, $5.8\%$ of participating pairs managed to login successfully with our threshold of $7$ correct answers for a successful authentication.  We conducted an analysis on the data to determine the true positive rate and the false positive rate in the form of a Receiving Operator Characteristics (ROC) graph (see Fig.~\ref{fig:ROCPNG}). According to the ROC graph, the resilience to the known adversary depends strongly on the threshold. At a threshold of $7$/$10$ correct questions, both false positive rate (FPR) and true positive rate (TPR) are low at $5.8\%$ and $11.7\%$, respectively (see Section \ref{subsection:usabilityanalysisandresults} for relevant discussion on usability analysis). The TPR can improve to $32.8\%$ by setting the threshold to $6$ at the cost of increasing FPR to $14.7\%$. However, $14.7\%$ is much higher than $1\%$ threshold, required for resilient to the known adversary. This analysis suggests that GeoSQ is not resilient to the known adversary.

\vskip 8pt
\noindent \textbf{Phishing.} Resilience to phishing attacks was identified as pertinent by the Bonneau \emph{et al.} framework \cite{Bonneau2012}. Classical phishing attacks work by tricking the user into entering their preset authentication credentials into a fraudulent site. Since GeoSQ's location questions would change over time, and will be used only once, a classical phishing attack would not work. This is because the attacker can't directly determine what location questions will be generated. Thus, we categorize GeoSQ as resilient to phishing attacks. 

\subsection{Usability Analysis and Results}
\label{subsection:usabilityanalysisandresults}

We will evaluate GeoSQ under core usability metrics, identified by Bonneau \emph{et al.}~\cite{Bonneau2012}, including efficiency of use, frequency of errors, and ease of learning.

\vskip 8pt
\noindent \textbf{Efficiency of Use.} An authentication method is \emph{efficient to use} if the time spent for each authentication is acceptably short, and a user can also set up his/her credentials within a reasonable time determined based on the target environment \cite{Bonneau2012}. For GeoSQ, the target environment is fallback authentication, therefore the comparison counterparts are commonly utilized fallback authentication methods such as security questions, email resets, and SMS resets.

\begin{figure}[tb]
\begin{tabular}{cc}
 \includegraphics[width=0.5\textwidth]{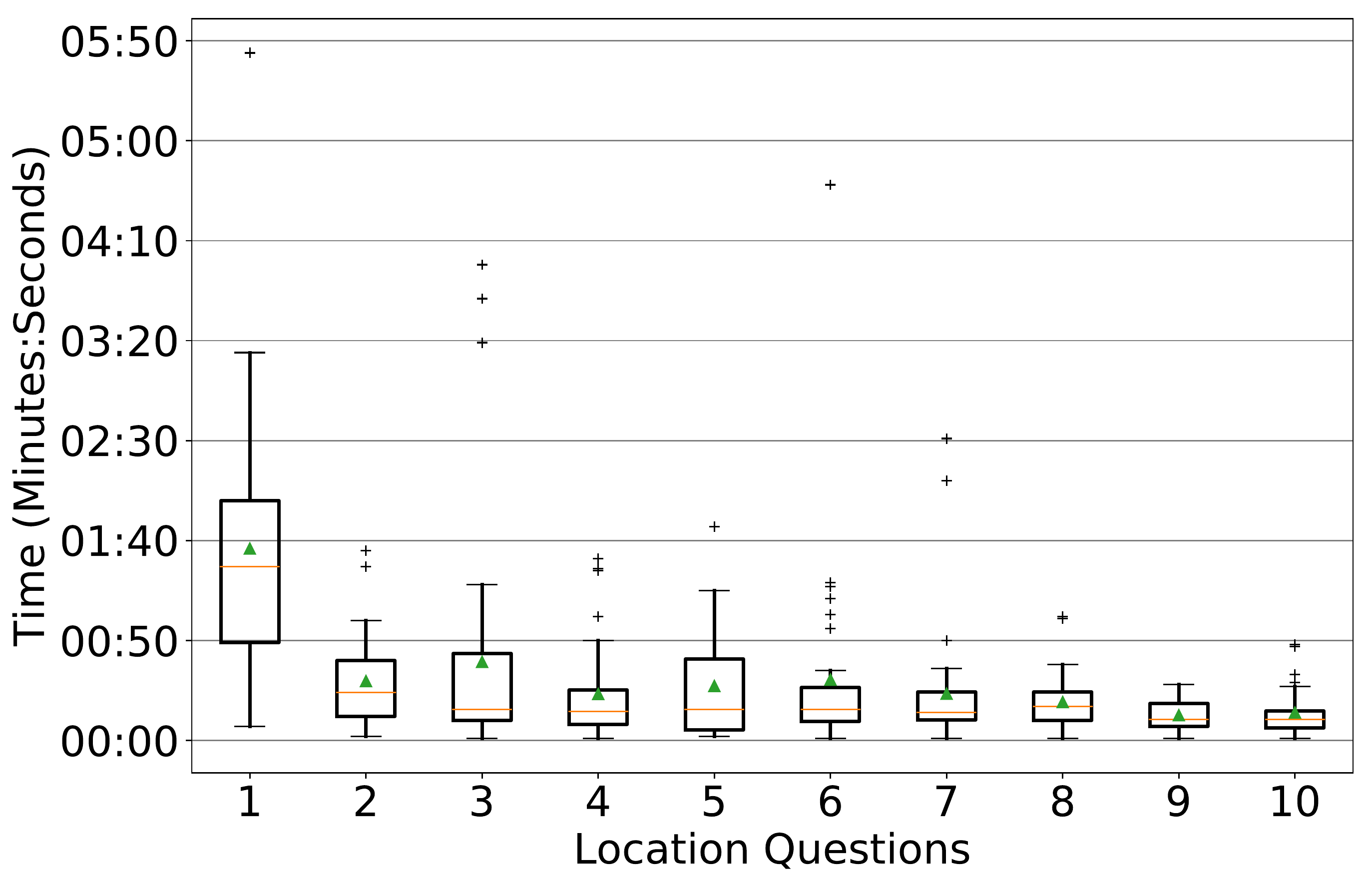} &  \includegraphics[width=0.5\textwidth]{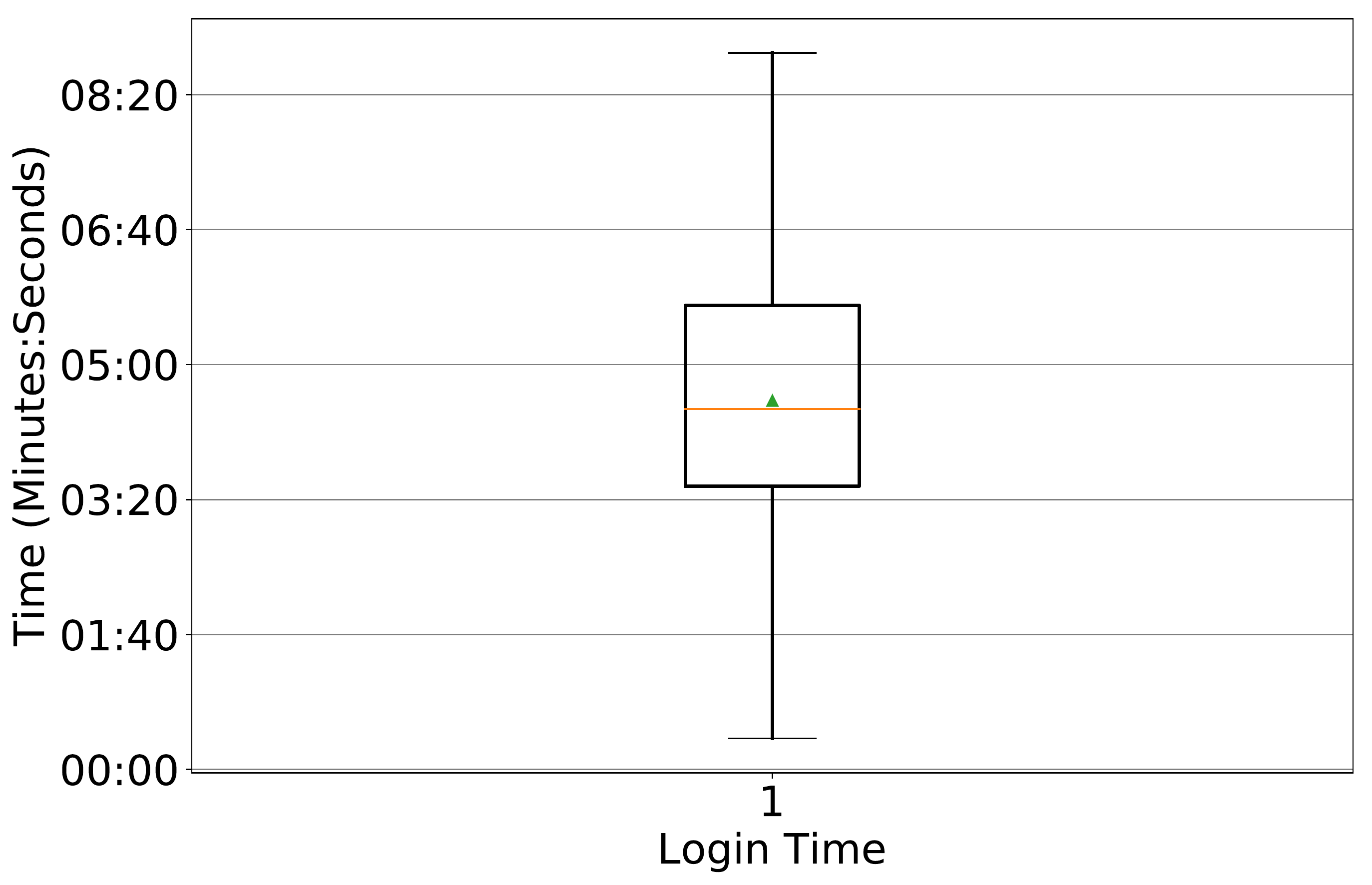}   \\
 
 (a) Login Time per Question  & (b) Total Login Time ($10$ questions)     \\
\end{tabular}

\caption{Login Time; (a) Average login time for GeoSQ for each question (n =$36$).  (b) Average login time of GeoSQ for all $10$ questions (n=$36$, each user was asked $10$ questions). Two outliers above 6 minutes were removed due to technical difficulties.}

\label{fig:logintimes}
\end{figure}

GeoSQ takes $7$-$11$ days to set up credentials (i.e., 10 unique locations) per use, unlike a security question, email reset, or SMS resets which take several seconds/minutes to set up fallback authentication credentials \cite{Bonneau2015}. This long set-up time of GeoSQ is due to its autobiographical nature. This usability flaw is mitigated by the fact that fallback authentication is not undergone as often as primary authentication by any typical user \cite{Bonneau2015}. Another important metric for efficiency of use is the login time. Fig.~\ref{fig:logintimes}a shows the average login time per question, and Fig.~ \ref{fig:logintimes}b shows the average login time for all ten questions in our user study.

\begin{figure}[tb]
\begin{tabular}{cc}
 \includegraphics[width=0.5\textwidth]{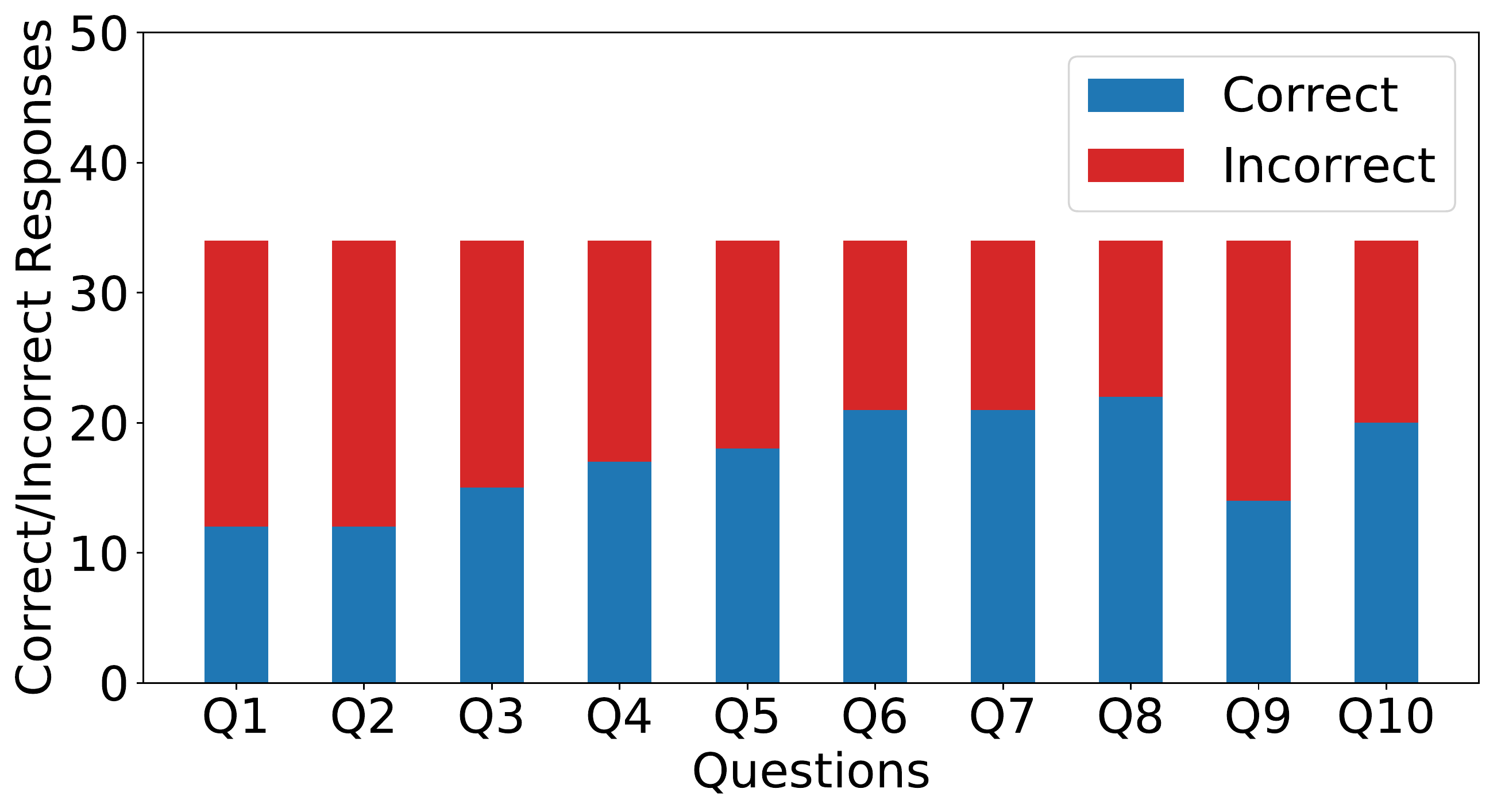} &  \includegraphics[width=0.5\textwidth]{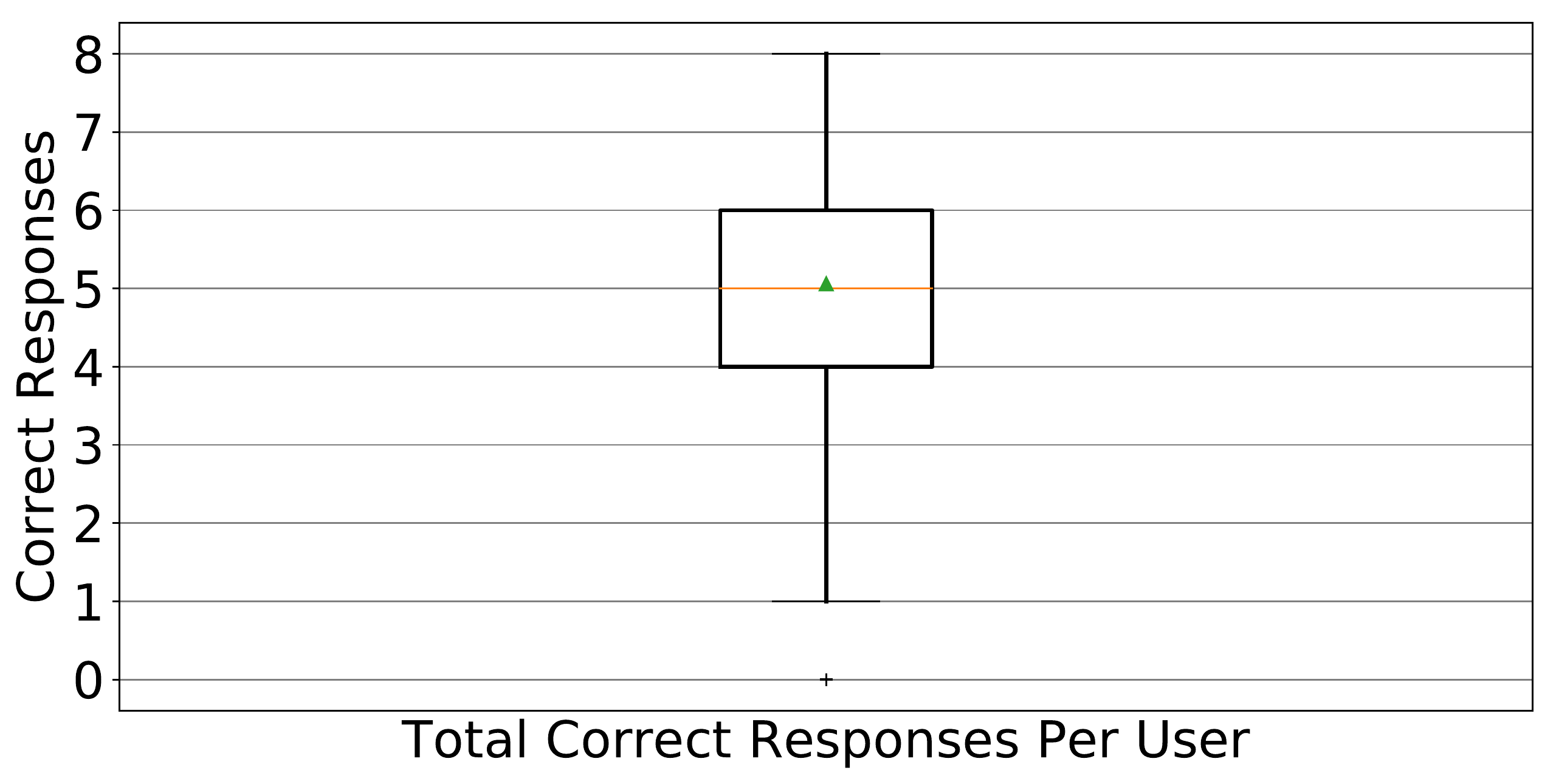}   \\
 
 (a) Correct/incorrect response per question  & (b) Average of correct responses \\
\end{tabular}

\caption{Correct/incorrect responses by legitimate users; (a) total number of correct and incorrect responses per question, by legitimate users (n=$36$, $360$ total attempts). (b) average correct responses (over questions and users)  (n=$36$).}

\label{fig:correctandincorrectresponses}
\end{figure}

The most noticeable outlier in Fig.~\ref{fig:logintimes}a is Q1 with an average higher than the other questions. We expected the login time of this question to be the lowest as it is about the most recent location.\footnote{GeoSQ orders the questions from the most recent location to the least recent one.} We further discuss about this outlier in Section \ref{subsec:user_study_observations}.  After question one, the average login time decreases with the number of location questions. This might suggest that the users are becoming more familiar with the interface. GeoSQ cannot be classified as efficient to use due to the long login time when compared to other commonly utilized fallback authentication methods.
\vskip 8pt
\noindent \textbf{Frequency of Errors.} An authentication scheme has infrequent errors if the login task is usually successful when performed by the true user \cite{Bonneau2012}. Fig.~\ref{fig:correctandincorrectresponses}a shows the number of correct/incorrect responses for all $36$ participants of Session $2$. Fig.~\ref{fig:correctandincorrectresponses}b shows the average correct responses over both users and questions. The average number of correct responses is $5.06$, slightly higher than $50\%$ of the questions asked. The participants were not allowed to attempt answering the same autobiographical location question if it was answered incorrectly and were simply moved to the next question.

Fig.~\ref{fig:ROCPNG} shows the ROC graph that determines the false positive rate and the true positive rate for each threshold. Given the number of incorrect answers in order to make this system usable for $61\%$ of users, we must set the threshold to $5$ correct answers out of $10$. However, the risk becomes the security concern because at that threshold, the false positive rate is about $24\%$ (n=$36$, $18$ pairs). 

Due to the high error rate of ~$50$\%, we consider GeoSQ highly prone to errors, requiring further investigation and improvement to make it usable.

\vskip 8pt
\noindent \textbf{Ease of Learning.} Any new authentication system has a learning curve. If an authentication system is too difficult for users to master, it becomes difficult to deploy it on a massive scale. 

To determine if the system was easy to learn and easy to use, we asked a series of usability questions in Session $2$ of our user study.  We asked two Likert Scale questions: (1) I thought the system was easy to use; (2) I thought the system was easy to learn to use. The users were asked to indicate the extent in which they agreed with those statements on a scale of $1$ to $7$.  

\begin{figure}[htbp!]
	\centering
 \includegraphics[width=\textwidth]{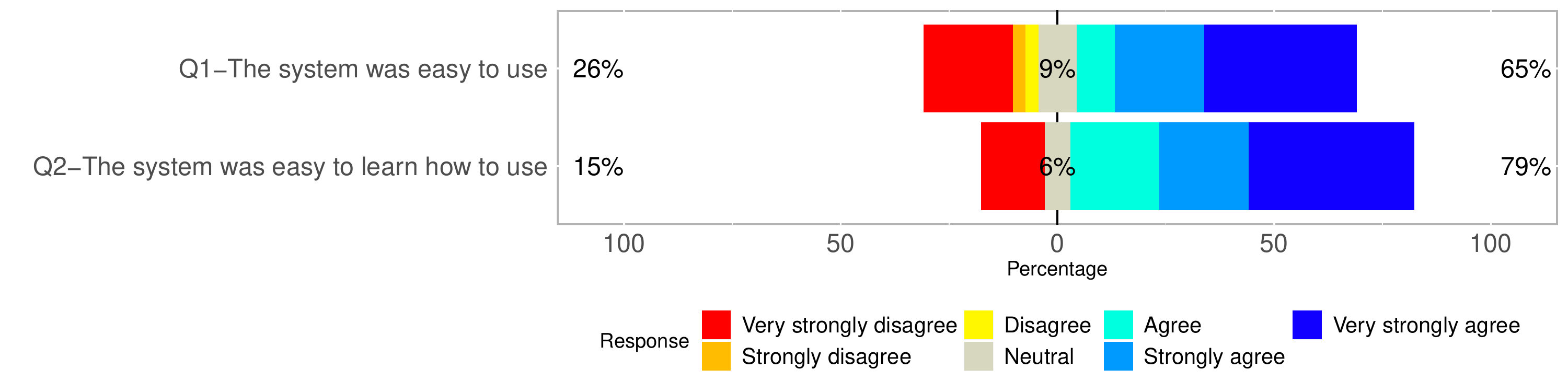}
	\caption{GeoSQ Usability Likert Scale questions}
	\label{fig:likert}
\end{figure}

Fig.~\ref{fig:likert} shows the responses to our questions: $65\%$ agreed, strongly agreed, or very strongly agreed that the GeoSQ system was easy to use; $11\%$ of our participants were neutral; and $26\%$  were either disagreed, strongly disagreed, or very strongly disagreed. For the second question, $79\%$ agreed, strongly agreed, or very strongly agreed that the system was easy to learn how to use. $6$\% were neutral and $26$\% were very strongly disagreed. Based on this, GeoSQ seems somewhat easy to use and easy to learn how to use.

\subsection{User Study Observations}
\label{subsec:user_study_observations}

We made several qualitative observations during the user study that can explain our results.

As participants were encouraged to use any online resource to aid in answering GeoSQ location questions, we noticed that most participants ($26/36 \sim 72\%$)  utilized their calendar on their laptop to find out what day of the week a certain date corresponded to. Our location questions were in the format of: $dd$ of $mm$ at $t$, where $dd$, $mm$, and $t$ stands for specific day, month, and time, respectively. We did not include what day of the week a certain date corresponded to in the location question. This lookup on calendar typically occurred for the first GeoSQ location question. This is a contributing factor to the long login time observed for Q1 (see Fig.~\ref{fig:logintimes}a). We believe including the day of the week in our location question could have improved the login time.   

Some participants also utilized their Google timelines\footnote{Google timelines is an online resource that tracks user locations and activities.}, and some others logged in to their transit applications to check their travel times. This also typically occurred for very first location question, which again has contributed to the long login time for Q1. 

Despite the demonstration of the GeoSQ interface at the start of Session $2$, the participants tended to fiddle around with the interface to get more acquainted with it. This is another contributing factor for the long login time of Q1 as noted in Section \ref{subsection:usabilityanalysisandresults}.

The default map view of GeoSQ is set to the current location. We observed that the participants were attempting to navigate by dragging the map instead of search functionality. As they became more acquainted to the interface, they used the search functionality. This possibly explains why the login time decreases as more location questions were asked (see Fig.~\ref{fig:logintimes}a).

\section{Discussion and Future Work}
\label{sec:Discussion}

GeoSQ's security is comparable to other fallback authentication techniques (See Appendix \ref{appendA} for evaluation under the Bonneau \emph{et al.} \cite{Bonneau2012} framework). However, GeoSQ's usability has several shortfalls, when compared to SMS resets, email resets, and security questions. 

The usability shortfalls arise from several design factors: (i) The time span to log $10$ unique questions was around $7$-$11$ days, that design decision was expected to harm usability to some extent. However, the long login time and the frequency of errors by participants originate from low memorability of autobiographical location information over $7$-$11$ days \cite{Albayram2016}, suggesting a need for decreasing the logging time-span. (ii) During Session $2$, we had a demonstration where GeoSQ features were explained. However, during the recall phase we observed many participants were still getting used to the interface because they did not have hands on experience. The lack of training was a contributing factor to the long login time which hindered usability. Hands-on training was required for better familiarity with the interface.  (iii) The absence of the day of the week in our autobiographical location questions turned out to be an important factor in slowing down the users. More informative questions by incorporating the day of the week could have potentially improve login time and memorability.

One can make a few conclusions with regard to GeoSQ's usability shortfalls: (i) From a memorability point of view, non-significant location events (e.g., going to a coffee shop) are expected to have weaker memorability compared to significant location events (e.g., attending a concert). However, there is no guarantee that each user has had significant location events in the recent past, and those events were logged for our users. Future work can study the detection of secure significant events for use in GeoSQ. (ii) GeoSQ can possibly be deployed for a subset of users. Individuals with location privacy concerns often keep their location services off. That's why we recommend using GeoSQ-like systems as an optional method for fallback authentication.\footnote{Providing options to users for fallback authentication is becoming common practice. Currently, users have choices between SMS resets, email resets, and security questions \cite{Bonneau2015}} (iii) Setting the error margin to be $200$ meters makes map based authentication more usable, and does not hinder the security to a great extent (see Section  \ref{subsec:securityanalysis}). Slightly increasing the errors margin from $200$ meters should improve the usability of GeoSQ.


GeoSQ offered strong security when compared to other fallback authentication methods. Its large key space ($2^{94}$ offers enough entropy to withstand throttled online guessing attacks. GeoSQ is also easy to learn how to use, despite the lack of training (relying only on a demonstration). We designed the system to closely resemble daily interactions with map based applications to make it easier for participants to become acquainted with the interface. However, the login time was relatively high as participants were getting acquainted with the interface.

Future work in location-based autobiographical authentication should focus on making questions more memorable by detecting significant events. Utilizing hints for location-based questions has proven to be effective in the past \cite{Albayram2015}. Hints paired with significant event detection has the potential to improve the memorability of similar authentication systems. Furthermore, current focus on location-based autobiographical authentication is on discrete location questions (i.e., where were you at a certain point in time). However, one can investigate location questions based on the sequence of locations or continuous locations (e.g., routes taken).

\section{Conclusion}
\label{sec:conclusion}

Fallback authentication techniques such as security questions, email resets, and SMS resets suffer from usability and security issues. The security of fallback authentication is very important, as one can bypass the primary method of authentication through them. In an attempt to address these issues, we developed GeoSQ  that logs location data, then asks $10$ unique location questions where $7$ correct answers are required for successful authentication. To examine GeoSQ for security and usability, we conducted a two-session in-person user study  (n=$19$ pairs). In this user study, we also asked pairs to guess each other's answers to evaluate the security of GoeSQ against the known adversary threat. Our results showed that GeoSQ is resilient against throttled guessing attacks due to its large key space of $2^{94}$. For the known adversary threat, adversaries obtained an average guessing score of ~$35$\% ($3.5$/$10$), with ~$5.8$\% of adversaries exceeding our $70$\% threshold for a successful authentication. We also found out that on average $50$\% of location questions were answered correctly by participants, with only about $11.7$\% of participants exceeding our $70$\% threshold for a successful authentication. This result indicates that utilizing a series of unique location based autobiographical questions that are unique is unusable due to the low memorability of episodic memory over longer time spans \cite{Conway2009}.   While GeoSQ does offer some security benefits (e.g., large key space) over currently utilized fallback authentication methods, improvements need to be made for its usability. Future work should examine improving the memorability of location based autobiographical authentication by detecting and utilizing significant events with potential for higher memorability. In addition, one can envision  investigating the utilization of continuous locations versus discrete locations (e.g., routes from point A to point B, versus asking about a specific location at a certain time).

\bibliographystyle{unsrt}  
\bibliography{references} 
\clearpage
\begin{appendices}
\section{}
\label{appendA}

\newcommand{\FC}{
	\begin{tikzpicture}
	\vspace{1em}
	\fill [ pattern= vertical lines, pattern color=white!60](-0.2,-0.2) rectangle (0.2,0.2);
	\draw [black,fill=black](0,0) circle (.8ex);
	\end{tikzpicture}
}

\newcommand{\EC}{
	\begin{tikzpicture}
	\vspace{1em}
	\fill [ pattern= vertical lines, pattern color=white!60](-0.2,-0.2) rectangle (0.2,0.2);
	\draw [black,fill=white](0,0) circle (.8ex);
	\end{tikzpicture}
}

\newcommand{\picQG}{
	\begin{tikzpicture}
	\vspace{1em}
	\fill [pattern= vertical lines, pattern color=green!60](-0.2,-0.2) rectangle (0.2,0.2);
	\draw [black,fill=white](0,0) circle (.8ex);
	\end{tikzpicture}}

\newcommand{\picG}{
	\begin{tikzpicture}
	\vspace{1em}
	\fill [ pattern= vertical lines, pattern color=green!60](-0.2,-0.2) rectangle (0.2,0.2);
	\draw [black,fill=black](0,0) circle (.8ex);
	\end{tikzpicture}}

\newcommand{\picQR}{
	\begin{tikzpicture}
	\vspace{1em}
	\fill [pattern= horizontal lines, pattern color=red!60](-0.2,-0.2) rectangle (0.2,0.2);
	\draw [black,fill=white](0,0) circle (.8ex);
	\end{tikzpicture}}

\newcommand{\picR}{
	\begin{tikzpicture}
	\vspace{1em}
	\fill [pattern= horizontal lines, pattern color=red!60](-0.2,-0.2) rectangle (0.2,0.2);
	\draw [black,fill=black](0,0) circle (.8ex);
	\end{tikzpicture}}

\newcommand{\SR}{
	\begin{tikzpicture}
	\vspace{1em}
	\fill [pattern= horizontal lines, pattern color=red!60](-0.2,-0.2) rectangle (0.2,0.2);
	\end{tikzpicture}}

\newcommand{\SG}{
	\begin{tikzpicture}
	\vspace{1em}
	\fill [pattern= vertical lines, pattern color=green!60](-0.2,-0.2) rectangle (0.2,0.2);
	\end{tikzpicture}}

\newcommand{\FT}{
	\begin{tikzpicture}
	\vspace{1em}
	\draw[black,fill=black] (-0.13,0) -- (0.13,0) -- (0,0.3) -- cycle;
	\end{tikzpicture}}

\newcommand{\ET}{
	\begin{tikzpicture}
	\vspace{1em}
	\draw[black,fill=white] (-0.13,0) -- (0.13,0) -- (0,0.3) -- cycle;
	\end{tikzpicture}}

\newcolumntype{R}[1]{>{\raggedleft\let\newline\\\arraybackslash\hspace{0pt}}m{#1}}

	\begin{table}[htbp!]
\centering
		\hspace*{-0.8cm}
		\setlength\tabcolsep{.5pt}
		\resizebox{\textwidth}{!}{%
		\begin{tabular}{||c|| m{.5cm} m{.5cm} m{.5cm} m{.5cm} m{.5cm} m{.5cm}m{.5cm} m{.5cm}|| m{.5cm} m{.5cm} m{.5cm} m{.5cm} m{.5cm} m{.5cm}|| m{.5cm} m{.5cm} m{.5cm} m{.5cm} m{.5cm} m{.5cm} m{.5cm} m{.5cm} m{.5cm} m{.5cm} m{.5cm} m{.5cm} m{.5cm} ||}
			\hline \hline
			&\parbox[c]{.35cm}{\rotatebox[origin=c]{90}{Memorywise-Effortless}} 
			
			&{\rotatebox[origin=c]{90}{Scalable for Users}}
			
			&{\rotatebox[origin=c]{90}{Nothing to Carry}} 
			
			&{\rotatebox[origin=c]{90}{Physically Effortless}} 
			
			&{\rotatebox[origin=c]{90}{Easy to Learn}}
			
			&{\rotatebox[origin=c]{90}{Efficient to Use}}
			
			&{\rotatebox[origin=c]{90}{Infrequent Errors}}
			
			&{\rotatebox[origin=c]{90}{Easy Recovery from Loss}}
			
			&{\rotatebox[origin=c]{90}{Accessible}}
			
			&{\rotatebox[origin=c]{90}{Negligible  Cost per User}}
			
			&{\rotatebox[origin=c]{90}{Server Compatible}}&{\rotatebox[origin=c]{90}{Browser Compatible}}
			
			&{\rotatebox[origin=c]{90}{Mature}}
			
			&{\rotatebox[origin=c]{90}{Non-Proprietary}}
			
			&{\rotatebox[origin=c]{90}{Res. to Physical Observation}}
			
			&{\rotatebox[origin=c]{90}{Res. to Targeted Impersonation}}
			
			&{\rotatebox[origin=c]{90}{Res. to Throttled Guessing}}
			
			&{\rotatebox[origin=c]{90}{Res. to Unthrottled Guessing}}
			
			&{\rotatebox[origin=c]{90}{Res. to Internal Observation}}
			
			&{\rotatebox[origin=c]{90}{Res. to Leaks from Other Verifiers}}
			
			&{\rotatebox[origin=c]{90}{Res. to Phishing}}
			
			&{\rotatebox[origin=c]{90}{Res. to Theft}}
			
			&{\rotatebox[origin=c]{90}{No Trusted Third Party}}
			
			&{\rotatebox[origin=c]{90}{Requiring Explicit Consent}}
			
			&{\rotatebox[origin=c]{90}{Unlinkable}}
			
			&{\rotatebox[origin=c]{90}{Res. to Social Media Mining}}
			
			&{\rotatebox[origin=c]{90}{Res. to the Known Adversary}}
			
			\\ \hline \hline			
			\multicolumn{1}{||m{2cm}||}{Passwords (Baseline)} & & & \FC  &  & \FC  &\FC  &\EC   &\multicolumn{1}{m{.5cm}||}{\FC } 	
			
			& \FC   &\FC &\FC   & \FC  & \FC  & \multicolumn{1}{m{.5cm}||}{\FC }
						
			& &\EC  &&&& & &\FC &\FC &\FC &\FC & \FT &\multicolumn{1}{m{.5cm}||}{\FT} \\ \hline

			\multicolumn{1}{||m{2cm}||}{Security \hspace{1ex} Questions} &\picQG & & \FC  &  & \FC &\FC &\EC &\multicolumn{1}{m{.5cm}||}{\FC} 
			& \FC &\FC &  & \FC  & \FC & \multicolumn{1}{m{.5cm}||}{\FC}
			& \SR & \SR & & &  & \SR & &\FC &\FC &\FC &\picR & \SR &\multicolumn{1}{m{.5cm}||}{\SR} \\ \hline	
			
			\multicolumn{1}{||m{2cm}||}{Email Resets} &\FC &\FC & \FC  &   &\FC &\FC &\FC &\multicolumn{1}{m{.5cm}||}{\FC} 
			
			& \FC &\FC & & \FC  & \FC & \multicolumn{1}{m{.5cm}||}{\FC }
			
			& & \EC & \picR & & \SG & \SR & \SR &\FC & &\FC & & \ET &\multicolumn{1}{m{.5cm}||}{\ET} \\ \hline	
			
			\multicolumn{1}{||m{2cm}||}{SMS Resets} & \FC &\FC &  &   &\FC &\FC &\FC &\multicolumn{1}{m{.5cm}||}{\FC} 
			
			& \FC &\FC & & \FC  & \FC & \multicolumn{1}{m{.5cm}||}{\FC }
			
			& & \EC & \FC & \ET & \SG & \SR & \SR &\EC & &\FC & & \ET &\multicolumn{1}{m{.5cm}||}{\ET} \\ \hline	
				
			\multicolumn{1}{||m{2cm}||}{GeoSQ} &\SR &\picR & \picQR  & \SR   &\FC &\SR &\SR &\multicolumn{1}{m{.5cm}||}{\SR} 
			
			& \FC &\FC & & \FC  & & \multicolumn{1}{m{.5cm}||}{\FC }
			
			&\FC & \picQR & \picR & \FC &\SG & \FC & \FC &\picR &\EC &\FC & & \picR &\multicolumn{1}{m{.5cm}||}{\SR} \\ \hline					
			
			\hline \hline
		\end{tabular}%
	}
		\vspace{3em}
		\begin{tablenotes}
			\small
			\item \FC \hspace{0.05cm} \textendash \hspace{0.05cm} offers the benefit, \EC   \hspace{0.05cm} \textendash \hspace{0.05cm} almost offers the benefit, no circle  \hspace{0.05cm} \textendash \hspace{0.05cm}  does not offer the benefit, \FT  \hspace{0.05cm} \textendash \hspace{0.05cm} potential to have the benefit , \ET  \hspace{0.05cm} \textendash \hspace{0.05cm} potential to almost offer the benefit,\SG  \hspace{0.05cm} \textendash \hspace{0.05cm} better than passwords, \SR  \hspace{0.05cm} \textendash \hspace{0.05cm} worse than passwords
		\end{tablenotes}
		   \caption{Adapted Bonneau \emph{et al.} framework \cite{Bonneau2012}. Analysis of Security Questions was conducted by Schechter \emph{et al.} \cite{Schechter2009}, analysis of web passwords was conducted by Bonneau \emph{et al.} \cite{Bonneau2010}. Please note that the two extra security categories that we added were not evaluated in the original framework proposal \cite{Bonneau2012}.}
  \label{tab:BonneauFramework}
	\end{table}

\end{appendices}

\end{document}